\documentclass[aps,prd,showpacs,twocolumn,showkeys,preprintnumbers]{revtex4}

\usepackage{graphicx}                                 
\usepackage{amsmath,amsfonts,mathbbol}
\graphicspath{{./Figures/}}                           

\newcommand{\ie}{\textit{\mbox{i.e.\ }}}              
\newcommand{\eg}{\textit{\mbox{e.g.\ }}}              
\renewcommand{\Re}{\mathfrak{Re}\,}                   
\newcommand{\Tr}{\mbox{Tr}}                           
\newcommand{\identity}{\mathbb{1}}
\newcommand{\IPR}{\mbox{\tt IPR}}                     
\newcommand{\bc}{\texttt{bc}}                         
\newcommand{\fc}{\texttt{fc}}                         
\newcommand{\Ncp}{N_{\rm cp}{}}                       
\newcommand{\Fig}[1]{Fig.~\ref{#1}}
\newcommand{\Tab}[1]{Table~\ref{#1}}
\newcommand{\Sec}[1]{Sec.~\ref{#1}}
\newcommand{\Eq}[1]{Eq.~(\ref{#1})}
\newcommand{\code}[1]{{\tt#1}}

\begin{document}


\preprint{HU--EP--05/63} 

\title{Spectral properties of the Landau gauge Faddeev-Popov operator
in lattice gluodynamics}

\author{A.~Sternbeck, E.-M.~Ilgenfritz, M.~M\"uller-Preussker}
\affiliation{Humboldt-Universit\"at zu Berlin, Institut f\"ur Physik,
  D-12489 Berlin, Germany} 

\date{September 30, 2005}

\begin{abstract}
  Recently we reported on the infrared behavior of the Landau gauge
  gluon and ghost dressing functions in $SU(3)$ Wilson lattice 
  gluodynamics with special emphasis on the Gribov problem. 
  Here we add an investigation of the spectral properties of the 
  Faddeev-Popov operator at $\beta=5.8$ and $6.2$ for lattice sizes 
  $12^4$, $16^4$ and $24^4$. The larger the volume the more 
  of its eigenvalues are found accumulated close to zero. Using 
  the eigenmodes for the spectral representation it turns out
  that for our smallest lattice $\mathcal{O}(200)$ 
  eigenmodes are sufficient to saturate the ghost propagator at lowest 
  momentum. We associate exceptionally large values of the ghost
  propagator to extraordinary contributions of low-lying eigenmodes.
\end{abstract}

\maketitle
%
\section{Introduction}
\label{sec:intro}

The infrared suppression of the gluon propagator on the one hand and
the enhancement of the ghost propagator at low momentum on the other
are closely related to the Gribov-Zwanziger horizon condition
\cite{Zwanziger:2003cf,Zwanziger:1993dh,Gribov:1977wm} as well as to
the Kugo-Ojima confinement criterion \cite{Kugo:1979gm}.
Zwanziger~\cite{Zwanziger:2003cf} has worked out that the continuum
behavior of both propagators in Landau gauge are consequences of
restricting gauge fields to the Gribov region $\Omega$, where the
Faddeev-Popov operator is non-negative. In general, for a given gauge
field a gauge orbit has more than one intersection (Gribov copy)
within $\Omega$. However, in the infinite volume limit expectation
values taken over arbitrary representatives of $\Omega$ are predicted
to be equal to those over the fundamental modular region $\Lambda$,
the set of gauge fields being absolute
maxima of the Landau gauge functional defined below. On 
a finite lattice, however, this equality cannot be
expected~\cite{Zwanziger:2003cf}. Therefore, the Gribov ambiguity has
to be explored in detail on finite lattices before drawing any
conclusion about the infrared behavior of the propagators
mentioned. In previous
investigations \cite{Cucchieri:1997dx,Bakeev:2003rr,Sternbeck:2005tk}
the influence of Gribov copies on the Landau gauge ghost and gluon
propagators was studied in detail both for $SU(2)$ and
$SU(3)$. Whereas the gluon propagator was not seen to be influenced by
the choice of gauge copies the ghost propagator turned out to be
clearly copy-dependent in the limit of small momenta.

In the present letter we are asking, how the singular behavior of the ghost 
propagator is related to the spectrum of the Landau gauge
Faddeev-Popov (F-P) operator. It is exactly there that the Gribov
ambiguity must become visible. We will demonstrate 
that the low-lying eigenvalues move towards zero as the volume
increases and that their values are sensitive to the choice of Gribov
copies. We will also discuss the localization properties of the corresponding
eigenmodes. The mode expansion of the momentum space ghost propagator will be
shown to converge the slower the higher the momentum is. 
Returning to the problem of exceptionally large values of the ghost 
propagator (``exceptional configurations''), discussed in
\cite{Bakeev:2003rr,Sternbeck:2005tk}, we can show that they are
related to strong contributions of the lowest eigenmodes. For the sake
of completeness we remind of a recent  
investigation of the spectrum of the Coulomb gauge F-P operator 
\cite{Greensite:2004ur}.

We introduce relevant definitions and notations in
\Sec{sec:definitions}. Simulation details are given in
\Sec{sec:simulation}. In \Sec{sec:FP-spectrum} the spectral properties
of the F-P operator are discussed. Configurations with exceptionally large 
values for the ghost propagator are studied 
in Section \ref{sec:exceptional}. In Section \ref{sec:conclusions} we
will draw our conclusions.

\section{Definitions}
\label{sec:definitions}

In order to study the ghost propagator using lattice simulations 
one has to fix the gauge for each thermalized $SU(3)$ gauge field
configuration $U\equiv\{U_{x,\mu}\}$. We apply the Landau gauge condition 
which can be implemented as a search for a gauge transformation 
$g \equiv \{g_x\}$
\begin{displaymath}
  U_{x,\mu} \rightarrow {}^{g}U_{x,\mu}=g_x\,
  U_{x,\mu}\,g^{\dagger}_{x+\hat{\mu}}\,, \quad g_x \in SU(3)
\end{displaymath}
that maximizes the Landau gauge functional
\begin{equation}
 F_{U}[g] = \frac{1}{4V}\sum_{x}\sum_{\mu=1}^{4}\Re\Tr \;{}^{g}U_{x,\mu}
 \label{eq:functional}
\end{equation}
while keeping the link variables $U_{x,\mu}$ fixed. 

The functional $F_{U}[g]$ has many different local maxima.
When the lattice volume $V$ is enlarged for a fixed inverse coupling
constant $\beta$ or, alternatively, $\beta$ is decreased, more and
more of these maxima become accessible by an iterative gauge fixing
process starting from various initial random gauge transformations. 
The set of gauge copies $\{{}^{g}U\}$ which correspond to different
(local) maxima of $F_{U}[g]$, where $U$ is kept fixed, are called
\emph{Gribov copies} in analogy to the Gribov ambiguity in the
continuum~\cite{Gribov:1977wm}. All Gribov copies   
belong to the gauge orbit created by $U$ and satisfy 
the differential (lattice) Landau gauge transversality condition
$(\partial_{\mu}{}^{g}\!\!A_{\mu})(x) = 0$ where 
\begin{equation}
  (\partial_{\mu}{}^{g}\!\!A_{\mu})(x) \equiv 
  \sum_{\mu} \left({}^{g}\!\!A_{\mu}( x+ \hat\mu/2)
                  -{}^{g}\!\!A_{\mu}( x- \hat\mu/2) \right) \, .
  \label{eq:transcondition}
\end{equation}
Here ${}^{g}\!\!A_\mu(x+\hat{\mu}/2)$ defines the non-Abelian
gauge potential on the lattice, \ie
\begin{equation}
  {}^{g}\!\!A_\mu(x+\hat{\mu}/2) \equiv \frac{1}{2i}\left(^{g} U_{x,\mu} -
    \ ^{g} U^{\dagger}_{x,\mu}\right)\Big|_{\rm traceless} \, .
\label{eq:A-definition}
\end{equation}
In the following, we will drop the label $g$ for convenience, \ie we assume 
$U$ to satisfy the Landau gauge condition such that $g\equiv\identity$
maximizes the functional in \Eq{eq:functional} relative to the
neighborhood of the identity.

The F-P operator is the Hessian of the gauge functional 
\Eq{eq:functional} and can be expressed in terms of the (gauge-fixed) link
variables $U_{x,\mu}$ as 
\begin{eqnarray}
  M^{ab}_{xy} & = & \sum_{\mu} A^{ab}_{x,\mu}~\delta_{x,y} 
  - B^{ab}_{x,\mu}~\delta_{x+\hat{\mu},y}
  - C^{ab}_{x,\mu}~\delta_{x-\hat{\mu},y}\quad
  \label{eq:FPoperator}
\end{eqnarray}
\begin{eqnarray*}
  \textrm{with}\quad
  A^{ab}_{x,\mu} &=& \phantom{2\cdot\ } \Re\Tr\left[ 
    \{T^a,T^b\}(U_{x,\mu}+U_{x-\hat{\mu},\mu}) \right],\\
  B^{ab}_{x,\mu} &=& 2\cdot\Re\Tr\left[ T^bT^a\, U_{x,\mu}\right],\\
  C^{ab}_{x,\mu} &=& 2\cdot\Re\Tr\left[ T^aT^b\, U_{x-\hat{\mu},\mu}\right]\;.
\end{eqnarray*}
Here $T^a$ ($a=1,\ldots,8$) denote the generators of the 
$\mathfrak{su}(3)$ Lie algebra satisfying $\Tr[T^aT^b]=\delta^{ab}/2$. 

For each maximum of the gauge
functional $F_U[g]$ the corresponding F-P operator has only positive
eigenvalues $\lambda_i > 0$, $(i=1,\ldots,8V-8)$, besides of its eight 
trivial zero modes, \ie the gauge-fixed field configurations $U$
lie within the Gribov region $\Omega$. We expect that the spectral
properties of the F-P operator differ for 
different Gribov copies. This should have consequences for the ghost
propagator. We will exploit the spectral
representation of the inverse of the F-P operator for a given gauge
field $U$ in terms of its real (ascendent) eigenvalues 
$\lambda_i$ and its (normalized) eigenvectors $\vec{\phi}_i(x)$ 
in coordinate space 
\begin{equation}
  [M^{-1}(U)]^{ab}_{xy} = 
  \sum_{i=1}^N \phi_i^a(x) \frac{1}{\lambda_i}
         \phi_i^b(y)\; .
\label{eq:Def-ghost-x-by-spectrum}
\end{equation}
The eigenvectors are given at each lattice point $x$ as
\mbox{$8$-component} color vectors $\vec{\phi}_i$ with components
$\phi^a_i(x)$. 
They are normalized such that $\sum_x |\vec{\phi}_i(x)|^2=1$.
Taking their Fourier transformed vectors $\vec{\Phi}_i(k)$ 
for momenta $k_{\mu} \in (-L_{\mu}/2, +L_{\mu}/2]$ and 
averaging over a Monte Carlo (MC) generated ensemble of gauge field
configurations we have computed the ghost propagator from
truncated mode expansions 
\begin{equation}
  G_n(q) = \langle G(k|n) \rangle_{\textrm{MC}} 
  \label{eq:Def-ghost-q-by-spectrum}
\end{equation}
where
\begin{equation} 
  G(k|n) = \frac{1}{8}~\sum_{i=1}^n
  \frac{1}{\lambda_i}\,\vec{\Phi}_i(k)\cdot\vec{\Phi}_i(-k)
  \label{eq:contribution}
\end{equation}
denotes the contribution of the eigenvalues and eigenmodes on a given
gauge field configuration. Here the vector and scalar product notation 
refers to the color indices. The Fourier momenta $k_{\mu}$ 
are related to the physical momenta $q_{\mu}$ by
\begin{displaymath}
 q_{\mu}(k_{\mu})=(2/a) \sin (\pi k_{\mu} / L_{\mu}) 
\end{displaymath}
with $a$ and $L_{\mu}$ denoting the lattice spacing and the linear
lattice extension, respectively.

If for the whole ensemble of configurations all \mbox{$N=8(V-1)$} non-trivial
eigenvalues and eigenvectors were known, the ghost 
propagator would be determined completely, \ie $G(q) \equiv G_N(q)$
at all momenta. We will check the convergence with respect to the order $n$  
by comparing the truncated propagator $G_n(q)$ with the full one, 
$G(q)$, obtained by inverting $M$ for 
a set of plane wave sources (with $\vec{k}\neq0$)
orthogonal to the trivial zero-modes. For details we refer to
reference \cite{Sternbeck:2005tk}.

From \Eq{eq:contribution} it is evident that the low-lying eigenvalues
and eigenvectors have a dominating impact on the ghost
propagator. However, for a finite lattice size it is not a priori
obvious what fraction of the F-P spectrum is responsible for the
enhancement of the ghost propagator at the smallest available momenta.
Zwanziger has argued that the F-P operator has very small eigenvalues
\cite{Zwanziger:2003cf}. In particular, it should have a high density
$\rho(\lambda)$ of eigenvalues per unit Euclicean volume at the Gribov
horizon (for $\lambda > 0$). This causes the ghost propagator to
diverge stronger than $1/q^2$ at $q=0$ \cite{Zwanziger:2003cf}. We estimate
the eigenvalue density $\rho$ at small $\lambda$ by 
\begin{equation}
  \label{eq:rho_lambda}
  \rho(\lambda) = \frac{h(\lambda, \lambda+\Delta\lambda)}{N \Delta\lambda}\, ,
\end{equation}
the average number $h$ of eigenvalues per gauge-fixed
configuration within the interval $[\lambda,\lambda+\Delta\lambda]$ 
divided by the bin size $\Delta\lambda$. For normalization the
denominator $N=8V$ has been chosen, 
since the F-P matrix is a $N \times N$ sparse symmetric matrix 
with $N$ linearly independent eigenstates. Note, the trivial zero modes 
are described by a $8\delta(\lambda)$-peak at $\lambda=0$.

\section{Simulation details}
\label{sec:simulation}

For the purpose of this study we have analyzed pure $SU(3)$ gauge
configurations thermalized with the standard Wilson action at two 
values of the inverse coupling constant $\beta=5.8$ and $6.2$. 
Cycles of one heatbath and four micro-canonical over-relaxation steps
were used for thermalization. As lattice sizes we used $12^4$, $16^4$ and
$24^4$. To each thermalized configuration $U$ a set of $\Ncp$ random
gauge transformations was assigned. Each has served as a starting
point for a gauge fixing procedure. We have applied standard
\emph{over-relaxation} with over-relaxation parameter tuned to
$w=1.63$. Keeping $U$ fixed this iterative procedure generates a
sequence of gauge transformations with increasing values of the gauge
functional (\Eq{eq:functional}). Thus, the final Landau gauge is iteratively 
approached until the stopping criterion in terms of the transversality 
(see \Eq{eq:transcondition})
\begin{equation}
  \max_x \Tr\left[\partial_{\mu} {}^g\!\! A_{\mu}(x)
      \partial_{\mu} {}^g\!\! A^{\dagger}_{\mu}(x)\right] < 10^{-14}
  \label{eq:stop_crit}
\end{equation}
was fulfilled. Consequently, each random start leads to its own local
maximum of the gauge functional. However, certain extrema of the
functional are found multiple times. This happened frequently
for the $12^4$ lattices, but rather seldom on larger ones. Note, we
used the worst local violation of transversality as stopping criteria 
which at a first glance seems to be very conservative. However, we
have found that the precision of transversality is crucial for the
final precision of the ghost propagator at low momentum.

In order to study the dependence on Gribov copies, for each $U$ 
that copy with largest functional value --- among all
$\Ncp$ gauge-fixed ones --- was stored, labeled as \emph{best copy} (\bc).
The larger $\Ncp$, the bigger the likeliness that this copy
represents the absolute maximum of the 
functional in \Eq{eq:functional}, at least with respect to the
observable in question. Indeed, with an increasing number $\Ncp$ of
inspected gauge copies for each $U$, the expectation values of the
ghost and gluon propagator evaluated on \bc{} copies is found to
converge more or less rapidly \cite{Sternbeck:2005tk}.
The first gauge copy was also stored, labeled as \emph{first copy} (\fc).
Obviously, this copy is as good as any other arbitrarily selected one.

On those ensembles of \fc{} and \bc{} gauge-fixed configurations 
the low-lying eigenvalues $\lambda$ of the \mbox{F-P}
operator and the corresponding eigenmodes have been separately
extracted, where we used the parallelized version of the
\code{ARPACK} package \cite{arpack}, \code{PARPACK}. To be specific,
the 200 lowest (non-trival) eigenvalues and 
their corresponding eigenfunctions have been calculated at $\beta=6.2$
using the lattice sizes $12^4$ and $16^4$ (see
\Tab{tab:statistics}). On the $24^4$ lattice, due 
to restricted amount of computing time, only 50 eigenvalues and
eigenmodes have been extracted at the same $\beta$.
In addition, 90 eigenvalues have been calculated on a $24^4$ lattice 
at $\beta=5.8$ providing us with an even larger physical volume. This
allows us to check whether low-lying eigenvalues are shifted towards
$\lambda \to 0$ as the physical volume is increased.

\begin{table}[htb]
  \centering
  \begin{tabular}{c@{\quad}cc@{\quad}c@{\quad}c@{\qquad}c}
    No. & $\beta$ & lattice & \# conf & \# copies & \# eigenvalues\\
\hline\hline
    1 & 6.2  & $12^4$  & 150 & 20 & 200\\
    2 & 6.2  & $16^4$  & 100 & 30 & 200\\
    3 & 6.2  & $24^4$  & 35  & 30 & 50\\
    4 & 5.8  & $24^4$  & 25  & 40 & 90\\
\hline\hline
  \end{tabular}
\caption{Statistics of the data used in our analysis. The last
  column lists the number of eigenvalues extracted separately on \fc{}
  and \bc{} copies of $U$. At $\beta=6.2$ the corresponding eigenmodes
  were calculated, too.}  
  \label{tab:statistics}
\end{table}

\section{The spectral properties of the F-P~operator}
\label{sec:FP-spectrum}

\subsection{The spectrum of the low-lying eigenvalues}

\begin{figure}[t]
  \includegraphics[width=8cm]{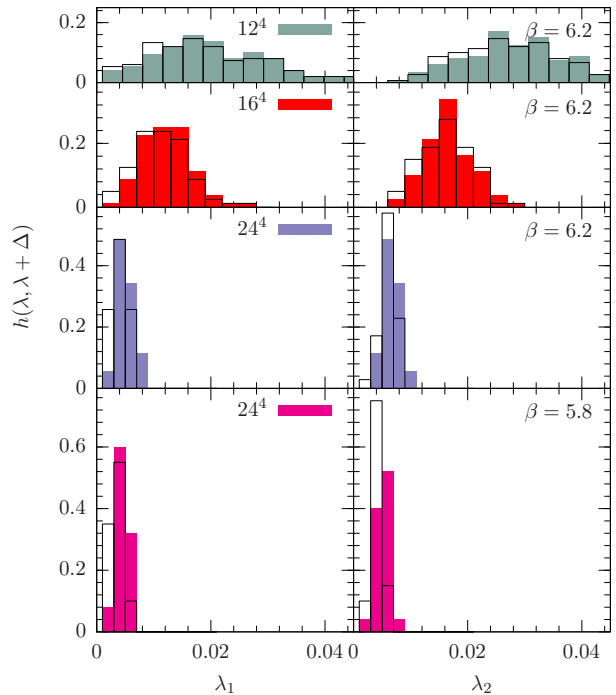}
  \caption{The frequency $h(\lambda)$ per configuration of the 
    lowest (left panels) and second lowest (right panels) 
    eigenvalue $\lambda$ of the Faddeev-Popov operator 
    is shown. Filled boxes represent the distribution obtained 
    on \bc{} gauge copies, while open ones represent 
    those on \fc{} copies.} 
  \label{fig:fps_lowest}
\end{figure}

Let us first discuss the distributions of the lowest $\lambda_1$ and
second lowest $\lambda_2$ eigenvalue of the \mbox{F-P} operator. These
are shown for different volumes in \Fig{fig:fps_lowest}. There
$h(\lambda,\lambda+\Delta)$ refers to the average number (per
configuration) of eigenvalues found in the intervall $[\lambda,\lambda+\Delta]$.
Open (full) bars refer to the distribution on \fc{} (\bc{}) gauge copies. 

\begin{figure}[htb]
  \centering \includegraphics[width=8.5cm]{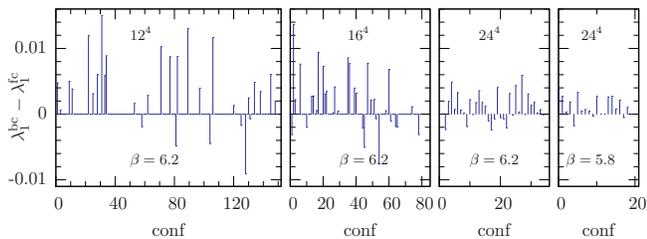}
  \caption{The differences $\lambda^{\bc}_1-\lambda^{\fc}_1$ 
    of the lowest F-P eigenvalues calculated on \bc{} and \fc{} 
    representatives for each gauge configuration are shown. 
    From left to right the lattice sizes are $12^4$, $16^4$ and 
    $24^4$ at $\beta=6.2$ and $24^4$ at $\beta=5.8$.}
  \label{fig:diff_lambda}
\end{figure}

From \Fig{fig:fps_lowest} it is quite obvious that both
eigenvalues $\lambda_1$ and $\lambda_2$ are shifted to lower values as
the physical volume is increased. In conjunction the spread of $\lambda$
values is decreased. This would be even more obvious, if we had shown
both distributions as functions of $\lambda$ in physical units.
It is also visible that the two low-lying  
eigenvalues $\lambda^{\fc}_i$ ($i=1,2$) on \fc{} gauge copies 
tend to be lower than those on \bc{}
copies. However, this holds {\it only on average} as can be seen from
\Fig{fig:diff_lambda}. There the differences
$\lambda^{\bc}_1-\lambda^{\fc}_1$ of the lowest eigenvalues 
on \fc{} and \bc{} gauge copies are shown for different lattice sizes
at $\beta=6.2$ and $5.8$. It is quite evident that there are few
cases where \mbox{$\lambda^{\bc}_1<\lambda^{\fc}_1$}, even though 
$F^{\bc}\ge F^{\fc}$ always holds for the gauge functional.

In addition we have checked how the average values $\langle\lambda\rangle$
of the eigenvalue distributions tends towards zero as the linear
extension $aL$ of the physical volume is increased. As in our previous study
\cite{Sternbeck:2005tk} we followed 
Ref.~\cite{Necco:2001xg} to fix the lattice spacing~$a$. For $\beta=5.8$
and 6.2 we used $a^{-1}$=1.446~GeV and 2.914~GeV, respectively, using the
Sommer scale $r_0=0.5$~fm. 

\begin{figure}[b]
  \includegraphics[height=7cm]{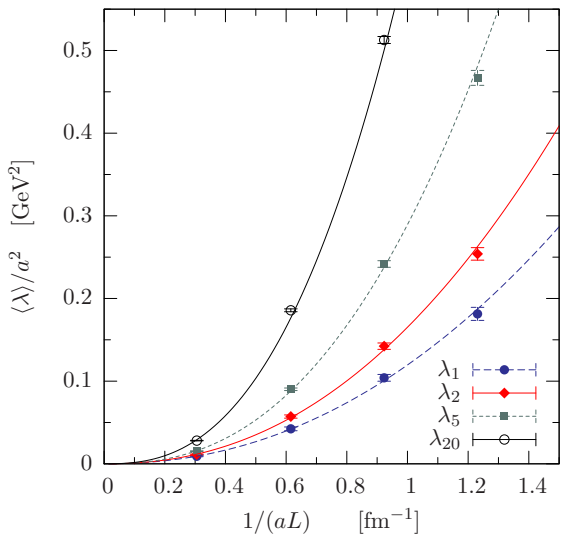}
  \caption{The average values $\langle\lambda_i\rangle/a^2$ (scaled to
    physical units) of the eigenvalues $\lambda_i$
    $(i=1,2,5,20)$ are shown vs. the inverse of the linear lattice
    extension $aL$. Only eigenvalues on
    \bc{} copies are shown. The lines represent fits
    to the data using the ansatz $a^{-2}\langle\lambda_i\rangle =
    C_i/(aL)^{2+\epsilon_i}$.}
  \label{fig:fps_mean}
\end{figure}

If the low-lying eigenvalues are supplemented with physical units 
it turns out that the average values of their distributions tend
towards zero stronger than $1/(aL)^{2}$. In fact, using the ansatz 
\begin{equation}
 f(aL) = \frac{C}{(aL)^{2+\varepsilon}} 
 \label{eq:lambda_aL_ansatz}
\end{equation}
to fit the data of $\langle\lambda_i\rangle/a^2$
for different $(aL)$, a positive $\varepsilon$ is found. The 
parameter of these fits are given in \Tab{tab:fps_fit} and in
\Fig{fig:fps_mean} we show the data and the corresponding fitting
functions.  There one clearly sees, the
low-lying eigenvalues not only approach zero, but also become closer
to each other with increasing $aL$. We have addressed the latter issue 
by fitting the differences
$(\langle\lambda_{i+1}\rangle-\langle\lambda_i\rangle)/a^2$ of adjacent
average values using the same  
ansatz \Eq{eq:lambda_aL_ansatz}. In \Tab{tab:fps_fit} we give the
parameter of those fits.

\begin{figure}[b]
  \includegraphics[height=7cm]{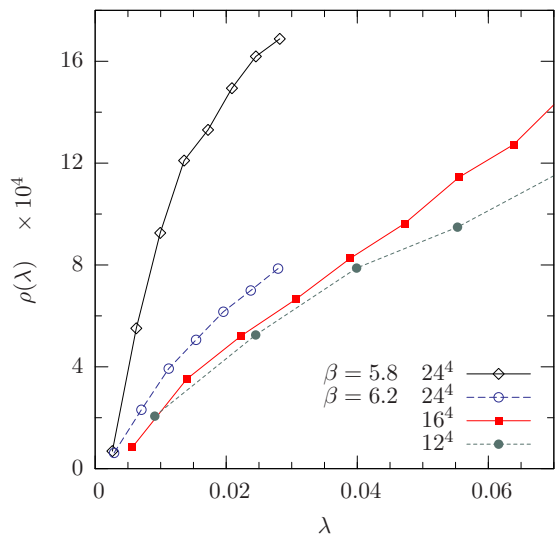}
  \caption{The eigenvalue density $\rho$ for  \bc{} copies 
    as a function of $\lambda$ estimated on $12^4$ and $16^4$ 
    lattices at $\beta=6.2$ and on a $24^4$ lattice for $\beta=6.2$ 
    and $5.8$. Bin sizes have been chosen as small as possible for each
    lattice size. The points marks the middle of each bin and the
    lines are to guide the eye.}
  \label{fig:fps_rho}
\end{figure}

As mentioned in \Sec{sec:definitions} the eigenvalue density $\rho(\lambda)$ 
is of particular interest. We have estimated this quantity according
to \Eq{eq:rho_lambda} where the bin sizes have been reasonably adjusted for 
the different volumes. In \Fig{fig:fps_rho} the estimates are shown for two 
values of $\beta$. There one clearly sees the eigenvalue
density close to $\lambda=0$ becomes a steeper function of $\lambda$ as
the physical volume becomes larger. 
It is remarkable that the increase
going from $\beta=6.2$ to $\beta=5.8$ on a $24^4$ lattice is even larger
than going from $12^4$ to $24^4$ at $\beta=6.2$, although in both cases
the physical volume is increased by a factor of about 16.

\begin{table}[t]
  \centering
  \begin{tabular}{c@{\qquad}c@{\quad}c@{\qquad}c}
\hline\hline
     $a^2f(aL)$    & $C$ & $\epsilon$ & $\chi^2/\textsc{ndf}$\\
\hline
    $\langle\lambda_1\rangle$ & 0.120(3) & 0.16(4) & 0.7\\
    $\langle\lambda_2\rangle$ & 0.165(4) & 0.24(5) & 1.8\\
    $\langle\lambda_5\rangle$ & 0.290(1) & 0.45(4) & 3.5\\
    $\langle\lambda_2\rangle-\langle\lambda_1\rangle$ & 0.045(2) &
    0.47(9) & 0.4\\ 
    $\langle\lambda_3\rangle-\langle\lambda_2\rangle$ & 0.051(1) &
    0.88(8) & 0.2\\ 
    $\langle\lambda_4\rangle-\langle\lambda_3\rangle$ & 0.033(1) &
    0.62(33) & 2.0\\ 
    $\langle\lambda_{5}\rangle-\langle\lambda_4\rangle$ & 0.037(1) &
    0.89(1) & 0.003\\
\hline\hline
  \end{tabular}
  \caption{The parameter $C$ and $\epsilon$ from fitting either the
    averages $\langle\lambda_i\rangle/a^{2}$ or the differences
    of adjacent average values
    $\langle\lambda_{i+1}\rangle/a^{2}-\langle\lambda_i\rangle/a^{2}$  
    of the corresponding eigenvalue distributions to the ansatz
    $f(aL) = C_i/(aL)^{2+\epsilon_i}$.}  
  \label{tab:fps_fit}
\end{table}

\begin{figure*}
  \includegraphics[width=0.8\textwidth]{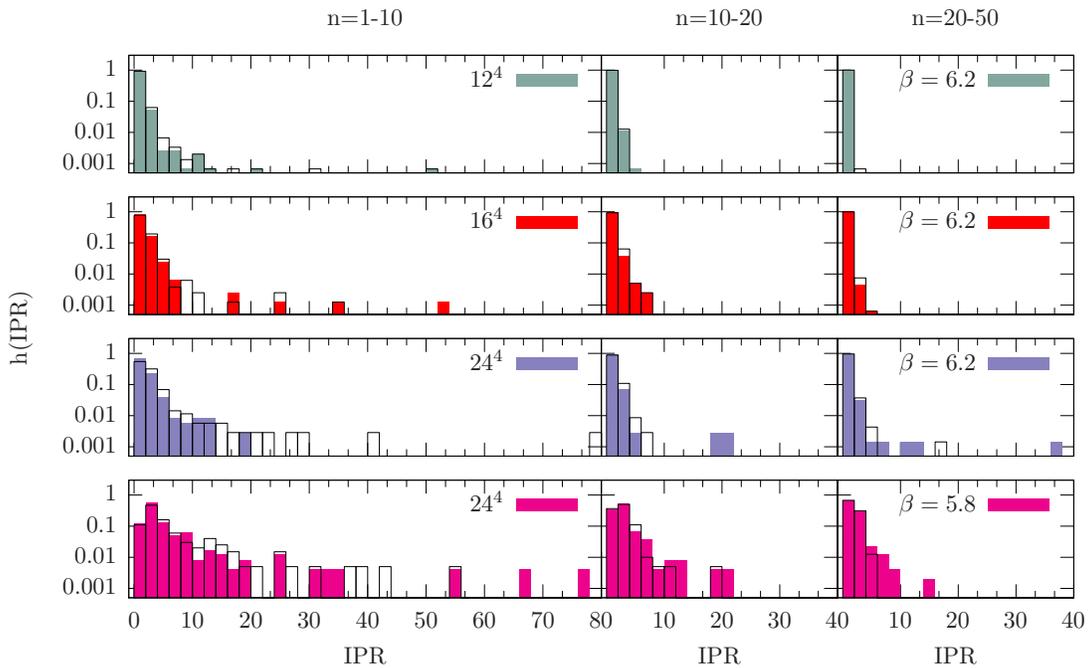}
  \caption{The relative distribution $h$ of \IPR{} values of the 10
    (left), the 10 to 20 (middle) and the 20 to 50 (right) lowest eigenmodes
    are shown. Note there is a logarithmic scale for $h(\IPR{})$. Each row
    corresponds to one pair of $\beta$ and lattice size. Filled boxes
    refer to distributions on \bc{} gauge copies, while open 
    ones correspond to \fc{} copies.} 
  \label{fig:ipr}
\end{figure*}

\subsection{Localization properties}

Together with the low-lying eigenvalues the
corresponding eigenvectors $\vec{\phi}(x)$ have been
extracted as well. It is interesting to evaluate for each vector
the inverse participation ratio (\IPR). The \IPR{} is defined as
\begin{displaymath}
   \IPR = V \sum_{x} |\vec{\phi}(x)|^4 \qquad\textrm{with}\; V=L^4
\end{displaymath}
and is a measure for the localization of an eigenvector. It enables
us to distinguish between eigenmodes with approximately uniformly
distributed values of $|\vec{\phi}(x)|^2$ \mbox{($\IPR\approx1\ldots
2$)} and specific ones with a small number of sites $x$ having large
modulus squared $|\vec{\phi}(x)|^2$ \mbox{($\IPR \sim
\rm{O}(100)$)}. Note, the eight (trivial) zero modes ($\lambda=0$) 
of the F-P operator give all $\IPR{}=1$. 

From \Fig{fig:ipr} we learn that the majority of eigenvectors of the
Faddeev-Popov operator is {\it not localized}. However, some large
$\IPR$ values have been found associated with modes among
the 10 lowest non-zero eigenmodes. This becomes more likely as the
volume is increased. So far we have no physical interpretation what
causes the stronger localization in these rare cases.

\subsection{What fraction of the F-P spectrum is dominating the ghost
  operator?}

As we have mentioned in \Sec{sec:definitions} there is 
an obvious way to construct the ghost
operator, if all eigenvalues $\lambda_i$ of the F-P operator and the
corresponding eigenvectors $\vec{\Phi}_i(k)$ in momentum space would
be available. However, their determination for each configuration is
numerically too demanding.

Restricting the sum in \Eq{eq:contribution} to the $n$ lowest
eigenvalues and eigenvectors ($n \ll N=8V-8$), we can figure out 
to what extent these modes, \ie the corresponding estimator $G_n(q)$ 
in \Eq{eq:Def-ghost-q-by-spectrum}, saturate the full ghost propagator
$G(q)$ obtained independently for a set of momenta by inverting the F-P
matrix on plane waves. See our recent study
\cite{Sternbeck:2005tk} for the data of $G(q)$.  

\begin{figure}[t]
  \includegraphics[width=8cm]{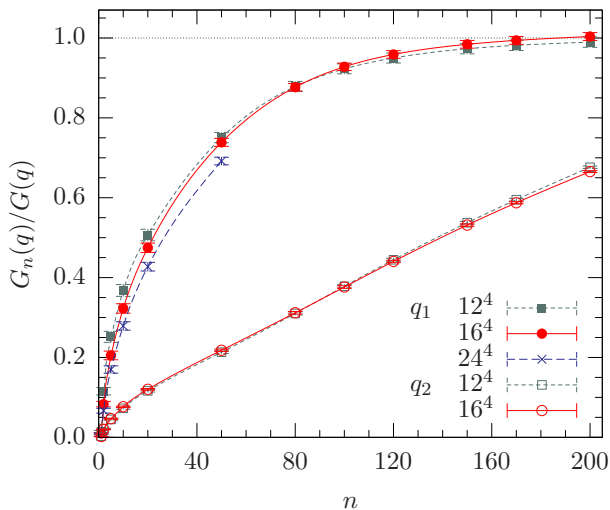}
  \caption{The ratio of the truncated ghost propagator $G_n(q)$ 
   (in terms of the $n$ lowest F-P eigenmodes and eigenvalues) to
   the full estimate $G(q)$ (taken from \cite{Sternbeck:2005tk}) 
   shown as a function of $n$ for the lowest ($q_1$) and second 
   lowest ($q_2$) momentum. The inverse coupling is $\beta=6.2$
   and the lattice size ranges from $12^4$ to $24^4$. 
   The data refer to \bc{} copies.}
  \label{fig:gh_vs_eigenmodes}
\end{figure}

This saturation is shown  in \Fig{fig:gh_vs_eigenmodes} for the lowest
$q_1$ and the second lowest momentum $q_2$ available on different
lattice sizes for $\beta=6.2$. There the values of $G_n(q)$ 
have been divided by the values for the full propagator $G(q)$ in
order to
compare the saturation for different volumes. Since $\vec{\Phi}_i(k)$
has been obtained by a fast Fourier transformation  
of the eigenvector $\vec{\phi}_i(x)$, all lattice momenta $k$
are available. Thus $G_n(q)$ refers to the average over all 
$k$ giving raise to the same momentum $q$. The full propagator values 
$G(q)$ at $q_1(k)$ and $q_2(k)$, however, refer to the averages over 
lattice momenta $k=([1,0],0,0)$ and to $k=(1,1,0,0)$, respectively. 

Let us consider first the lowest momentum $q_1$. We observe from
\Fig{fig:gh_vs_eigenmodes} that the approach to convergence differs,
albeit slightly, for the three different lattice sizes. The relative
deficit for $n<50$ rises with the lattice volume. For $n>100$ the rate
on a $16^4$ lattice is even a bit larger than that on a $12^4$
lattice. Unfortunately, there are no data available for $n>50$ on the
$24^4$ lattice. However, for the $12^4$ and $16^4$ lattices the rates
of convergence are about the same.  For example, taking 
only 20 eigenmodes one is definitely far from saturation (by about
50\%) whereas $150 \ldots 200$ eigenmodes are sufficient to reproduce
the ghost propagator within a few percent. In other words, the ghost
propagator at lowest momentum on a $12^4$ ($16^4$) 
lattice is formed by about 0.12\% (0.03\%) of the lowest
eigenvalues and eigenfunctions of the \mbox{F-P} operator.

For the second lowest momentum $q_2$ the contribution of even 200
eigenmodes is far from being sufficient to approximate the propagator.
%

\section{The problem of exceptional configurations}
\label{sec:exceptional}

We turn now to a peculiarity of the ghost propagator at larger $\beta$
of which we have reported in \cite{Sternbeck:2005tk}. It was also seen by
two of us in an earlier $SU(2)$ study \cite{Bakeev:2003rr}. While
inspecting our data we found, though rarely, that there are exceptionally 
large values in the Monte Carlo (MC) time histories of the ghost propagator 
at lowest momentum. Those values are not equally distributed around the 
average value, but rather are significantly larger. For details we refer 
to reference \cite{Sternbeck:2005tk}.

We have tried to find a correlation of such exceptionally large values in 
the history of the ghost propagator with other quantities measured
in our simulations. For example we have checked whether there is a direct
correlation between the values of the ghost propagator $G(k)$ as they
appear in the MC time histories (see \eg Fig.~5 in \cite{Sternbeck:2005tk})
and the lowest eigenvalue $\lambda_1$ of the F-P operator.

\begin{figure}[t]
  \includegraphics[width=0.45\textwidth]{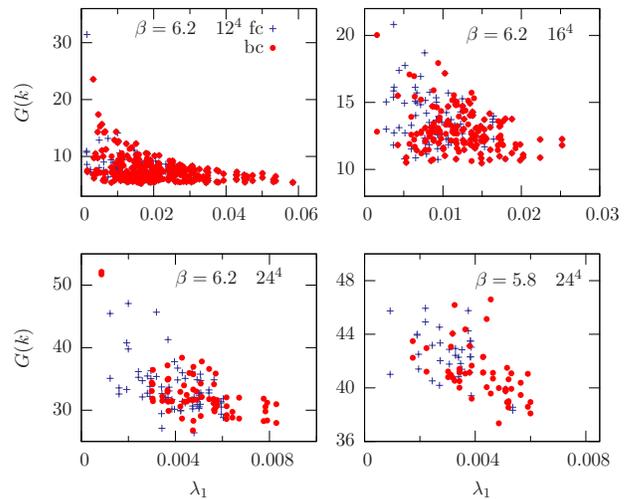}
  \caption{Scatter plots of MC time history values of the ghost
    propagator $G(k)$ at \mbox{$k=([1,0],0,0)$} vs.~the lowest F-P eigenvalue
    $\lambda_1$ are shown. The upper panels show data at $\beta=6.2$ on a $12^4$
    (left) and $16^4$ (right) lattice,  the lower ones on a $24^4$
    lattice at $\beta=6.2$ (left) and $\beta=5.8$ (right). Crosses
    refer to \fc{} gauge copies and filled circles to \bc{} copies.}
\label{fig:fps_gh}
\end{figure}

In \Fig{fig:fps_gh} we show such correlation in a scatter plot for
different lattice sizes at $\beta=5.8$ and $6.2$. There each entry
corresponds to a pair $[\lambda_1,G(k)]$ measured on a given gauge copy
of our sets of \fc{} and \bc{} copies. It is visible in that figure,
gauge copies giving rise to an extremely large MC value for the ghost
propagator are those with very low values for $\lambda_1$. This holds
for the $12^4$ and $16^4$ lattice. However, a very low eigenvalue is
not sufficient to obtain large MC values for $G(k)$ as can
be seen in the same figure. It is not excluded that such gauge copies
with extremely small eigenvalues would turn out to be
\emph{exceptional} for another realization of lowest  
momentum $q(k)$ than those two we have used. This might explain
why some configurations with extremely small lowest eigenvalues were not 
found to be exceptional with respect to the ghost propagator at $k=(1,0,0,0)$ 
and $k=(0,1,0,0)$. 

\begin{figure}[t]
\includegraphics[width=8cm]{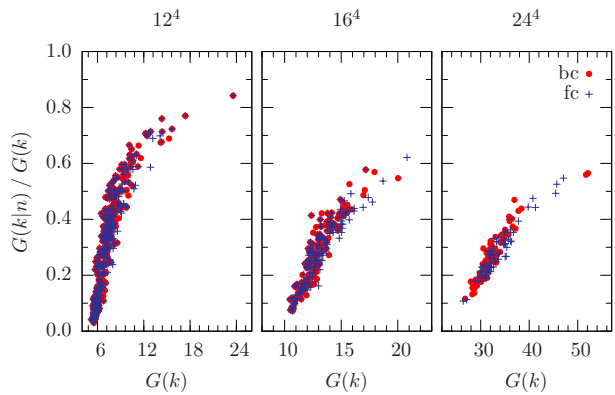}
\caption{Scatter plot of the relative contribution of the truncated sums 
  $G(k|n)$ over eigenmodes (see \Eq{eq:contribution}) to the full
  ghost propagator values $G(k)$ versus $G(k)$ for lattice momenta 
  \mbox{$k=(1,0,0,0)$} and \mbox{$k=(0,1,0,0)$}. From left to
  right the lattice sizes are $12^4$, $16^4$ and
  $24^4$ all for $\beta=6.2$. Data for \fc{} and \bc{} gauge copies 
  have been plotted separately.}
\label{fig:gh_vs_lowest_ev} 
\end{figure}

In the light of \Eq{eq:contribution} it is not adequate to
concentrate just on the lowest eigenvalues. Instead, one can monitor
the contribution of a certain number of eigenvalues $\lambda_i$ and 
eigenmodes $\vec{\Phi}_i(k)$ to the ghost propagator at some momentum
in question. Therefore, we have compared the truncated sums $G(k|n)$
according to \Eq{eq:contribution} 
with the MC history values of the full ghost 
propagator $G$. In fact, we show in the scatter plots in
\Fig{fig:gh_vs_lowest_ev} the ratios $G(k|n)/G(k)$ versus $G(k)$ for
$n=10$ and for various lattice sizes.

Obviously there is a strong correlation between the chosen group
of low-lying modes and the MC time history values of the full ghost propagator.
Indeed, if we consider values $G(k)>15$ to be \emph{exceptional} in the 
left-most panel ($12^4$ lattice) we find that the contribution of 
the 10 lowest modes amounts to more than 75\% of the actual value of 
the ghost propagator. On the opposite, for low $G(k)$ values the main 
contributions come necessarily from the higher eigenmodes, 
while the 10 lowest modes contribute a minor part only. A similar
but less dominant contribution of the 10 lowest modes is found for the 
time histories produced on larger lattices ($16^4,~24^4$). 

\section{Conclusions}
\label{sec:conclusions}

In this paper we have investigated the spectral properties of the F-P
operator and their relation to the ghost propagator in $SU(3)$ Landau
gauge. The configurations under examination have been generated on a $24^4$
lattice at $\beta=5.8$ and on $12^4$, $16^4$ and $24^4$ lattices 
at $\beta=6.2$.

As expected from Ref. \cite{Zwanziger:2003cf} we have found that 
the low-lying eigenvalues are shifted towards $\lambda=0$ as the
volume is increased. The result is an eigenvalue density
$\rho(\lambda)$ becoming steeper rising close to  
$\lambda=0$. We have also shown that the Gribov ambiguity is reflected
in the low-lying eigenvalue spectrum. In fact, the low-lying
eigenvalues extracted on \bc{} gauge copies are larger on average than
those on \fc{} copies. Thus better gauge-fixing (in terms of the gauge
functional) inhibits the above-mentioned tendency, keeping the gauge-fixed 
configuration slightly away from the Gribov horizon.

The study of the ghost propagator in terms of the
eigenvalues and eigenmodes of the F-P operator reveals that
there is a dominance of the low-lying part of the spectrum at
lowest momentum.
About 200 eigenmodes are sufficient to reconstruct the asymptotic
result up to a few percent at the lowest momentum on a $12^4$
lattice at $\beta=6.2$. With respect to the whole set of $8(12^4-1)$
non-trivial eigenvalues, this is a fraction of about 0.12\%. For
larger volumes the number of necessary  
eigenmodes seems to be somewhat larger. For the next higher momentum,
saturation needs a much bigger part of the low-lying spectrum.

On average the \mbox{F-P} eigenmodes are not localized, however, few 
large values have been seen among the lowest eigenmodes which 
so far could not be correlated to other measured quantities.

Analogously with observations made in Ref.~\cite{Bakeev:2003rr} 
we have reported recently \cite{Sternbeck:2005tk} on exceptionally 
large values in the Monte Carlo history of the ghost propagator. 
These we have seen at $\beta=6.2$ and only for some lattice
momenta $k$ realizing the lowest physical momentum $q(k)$. In the
study at hand we have shown that these \emph{outliers} can be assigned to the
contribution of the ten lowest \mbox{F-P} eigenmodes to the ghost
propagator at this particular $k$.

\section*{ACKNOWLEDGMENTS}

All simulations have been done on the IBM pSeries 690 at HLRN.  We
thank R.~Alkofer and L.~von Smekal for discussions. We are indebted 
to H.~St\"uben for contributing parts of the program code. 
A.~Sternbeck acknowledges support of the DFG-funded graduate school 
GK~271. This work has been supported by the DFG under contract
FOR~465 (Mu 932/2).  


\bibliographystyle{apsrev} \bibliography{references}

\end{document}